\documentclass[twocolumn,a4paper]{article}

\usepackage{bm} % bold math
\usepackage{amssymb,amsfonts}
\usepackage{abstract}
\usepackage{color}
\usepackage[pdftex]{hyperref}

\begin{document}

\title{\bf{\huge{Emergence of space-time and gravitation}}}
\author{\bf{Walter Smilga} \\ Isardamm 135d, 82538 Geretsried, Germany \\ Email:\,\,wsmilga@compuserve.com }

%\date{Received 11.02.2013}

\twocolumn[
  \begin{@twocolumnfalse}
    \maketitle
		
\begin{abstract}
In relativistic quantum mechanics, elementary particles are descri\-bed by 
irreducible unitary representations of the Poincar\'e group.
The same applies to the center-of-mass kinematics of a multi-particle system
that is not subject to external forces.
As shown in a previous article, for spin-1/2 particles, irreducibility leads 
to a correlation between the particles that has the structure of the 
electromagnetic interaction, as described by the perturbation algorithm of 
quantum electrodynamics.
The present article examines the consequences of irreducibility for a
multi-particle system of spinless particles.
In this case, irreducibility causes a gravitational force, which in the 
classical limit is described by the field equations of conformal gravity.
The strength of this force has the same order of magnitude as the strength 
of the empirical gravitational force.
\end{abstract}

{\bf Keywords:} conformal gravity; quantum gravity; emergence of space-time

\vspace{1cm}

\end{@twocolumnfalse}
]

\maketitle 

%{\bf PACS} 04.50.Kd, 04.60.-m, 11.30.Cp, 12.90.+b

\section{Introduction}

As a general rule of relativistic quantum mechanics, not only elementary 
particles, but also compound systems of particles are described by irreducible 
unitary representations of the Poincar\'e group, as long as no external 
forces act on the system.

Within a two-particle state, irreducibility of the representation that describes
the center-of-mass kinematics, causes a correlation of the individual particle 
momenta.
In a previous article \cite{ws}, the author has shown that for spin-1/2 particles, 
the quantum mechanical formulation of this correlation takes on the structure 
of the electromagnetic interaction, as described by the perturbation algorithm 
of quantum electrodynamics.
The coupling constant, derived from the geometrical properties of this correlation, 
was found to be in excellent agreement with the experimental value of 
the electromagnetic fine-structure constant.
This agreement emphasizes the crucial role of irreducibility for the 
kinematics of quantum mechanical multi-particle systems. 

Irreducible representations of the Poincar\'e group are labeled by the 
values of two Casimir operators $P$ and $W$ (see, e.g., \cite{sss})
\begin{equation}
P = p_\mu p^\mu,    \label{1-1}
\end{equation}
where $p_\mu$ is the total 4-momentum of the system,
and
\begin{eqnarray}
W = -w^\mu w_\mu , \; \mbox{with} \;\; 
w_\sigma = \frac{1}{2} \epsilon_{\sigma \mu \nu \lambda} 
M^{\mu \nu}p^\lambda ,   \label{1-2}
\end{eqnarray}
which refers to the angular momentum of the system.

Whereas the previous article was primarily based on the first Casimir 
operator $P$, the present article will concentrate on the second Casimir 
operator $W$. 
This operator is related to the intrinsic angular momentum of the two-particle
system, generated by the relative motion of the particles. 
 
Let $p_1$ and $p_2$ be the 4-momenta of two particles, for simplicity with 
equal masses $m$, so that
\begin{equation}
p = p_1 + p_2       \label{1-20}
\end{equation} 
denotes the total momentum and          
\begin{equation}
q = p_1 - p_2,        \label{1-21}
\end{equation}
the relative momentum.
Then $p$ and $q$ satisfy
\begin{equation}
p\,q = 0 .             \label{1-22}
\end{equation}
Based on Equation~(\ref{1-22}), a two-particle system can be described by a total 
momentum $p$ and a spacelike momentum $q$, perpendicular to the timelike 
vector $p$.
``Perpendicular to a timelike vector'' means that $q$ is allowed to rotate by 
the action of an SO(3) subgroup of the Lorentz group.
So the kinematics of the relative momentum is restricted to a 3-dimensional 
subspace of space-time.
 
For an irreducible two-particle representation, we obtain from the
constancy of the Casimir operator $P$ 
\begin{equation}
p^2 = (p_1 + p_2)^2 = M^2  , \label{1-3}
\end{equation}
where $M$ is the ``mass'' of the two-particle system, and
\begin{equation}
q^2 = 4m^2 - M^2 \le 0 , \label{1-24}
\end{equation} 
and further
\begin{equation}
2 p_1 p_2 = M^2 - 2m^2     \label{1-25}
\end{equation}
and
\begin{equation}
2 p_1 p = 2 p_2 p = M^2   .  \label{1-26}
\end{equation}
Equations~(\ref{1-25}) and (\ref{1-26}) correlate the particle momenta by 
fixing the angle between them and with respect to the total momentum $p$.
Rotations with rotational axis $p$ preserve these angles. 
Since these rotations leave $p$ invariant, they can be related to an
independent, internal degree of freedom, described by an action 
of SO(2) on the relative momentum $q$.

From the quantum mechanics of angular momentum, we know that 
for large quantum numbers the property of spherical symmetry does not
fade away, but is preserved, at least in the sense of an SO(2) symmetry:
for large quantum numbers, the spherical harmonics describe the orbits 
with a circular symmetry (cf.\,\,e.g.\,\,\cite{are}).
So, as in the foregoing quasi-classical consideration, we again encounter
circular orbits. 

This obviously means that in a two-particle state that is part of an irreducible 
two-particle representation of the Poincar\'e group, the individual particles 
are by no means in straight uniform motion.
According to Newton's first law \cite{in},
\textit{Corpus omne perseverare in statu suo quiescendi vel movendi uniformiter 
in directum, nisi quatenus a viribus impressis cogitur statum illum mutare}, 
they rather behave like two particles being forced into a circular orbit by a
kind of attractive gravitational force.

Such a force of obviously universal character has not been seen in the
experiments of particle physics---or perhaps, for some reason, it has been 
ignored.
This article is intended to find out more about this force, which obviously 
is the outcome of a combination of quantum mechanics and relativistic 
invariance.

\section{Parameter space-time vs.\\physical space-time}

Our analysis starts with a review of the role of space-time within the 
formalism of quantum dynamics.

Given an elementary particle, described by an irreducible representation of 
the Poincar\'e group in a state space with eigenstates $| p \rangle$ of the 
4-momentum $p$, then states ``in space-time'' can be defined by superposing 
these momentum eigenstates: 
\begin{equation}
| \mathbf{x},t \rangle = (2\pi)^{-\frac{3}{2}} \int \frac{d^3 
\mathbf{p}}{2|p_0|}\, e^{-\frac{ipx}{\hbar}} | p \rangle, 
\;\;k=1,2,3,    \label{2-1}
\end{equation}
with parameters $x = (\mathbf{x}, t)$. 
A detailed discussion of these states can be found in \cite{nw}.
See also \cite{rhbook}.

The parameters $x$ form a parameter space ($x$ space) with the same metric as 
the energy-momentum space ($p$ space).
The states $| \mathbf{x},t \rangle$ are ``localized'' (within a Compton wave length) 
at time $t=x_0$ at the point $\mathbf{x}$ of three-dimensional space.
So we can say that the $x$ space has also a ``physical'' meaning in the 
sense that it is a space in which (isolated) particles can be physically placed.

Note that Definition~(\ref{2-1}) does not require a prior existence of space-time.
It rather defines {\it space-time} on the basis of the momentum eigenstates.
We also define a \begin{it}position operator\end{it} in three-dimensional space by 
\begin{equation}
X_k = -i \hbar \frac{\partial}{\partial p^k}  . \label{2-2}
\end{equation}
The definition of a corresponding ``time'' operator does not make sense, because 
the states (\ref{2-1}) cannot be ``localized in time.''
Therefore, time is not an observable, but merely a parameter.
By Definition~(\ref{2-1}), space-time is derived as a property of matter, just as 
momentum is considered a property of matter.

The relation between $x$ space and $p$ space contains Planck's 
constant $h$.
This is the result of having independent scales for $x$ and $p$. 
We can avoid this constant by replacing $p$ by the {\it wave vector} $k$, 
defined by $p = \hbar\,k$, which in this context may be a more 
natural choice.

Now consider two elementary particles, described by an irreducible 
two-particle representation of the Poincar\'e group.
Because of the constraints from the two-particle mass shell relation 
Equation~(\ref{1-3}), it is not possible to simultaneously construct localized 
states for each particle.
Therefore, when two or more particles are considered, the physical property of 
$x$ space may be lost, but it still can serve as a useful parameter space, 
e.g., for {\it wave functions}. 
So we have to be careful not to mix up \begin{it}parameter\end{it} space-time 
with \begin{it}physical\end{it} space-time.
In the following, physical space-time will be understood in the sense of the 
{\it expectation value} of the position operator of Equation~(\ref{2-2}).

As a pure mathematical construct, (parameter) $x$ space is not limited by 
any ``physical'' scale, such as the Planck length.
So it does not make sense to try its ``quantization'' at Planck scales, in the 
hope of finding a road to quantum gravity. 
On the other hand, physical space-time is quantized right from the beginning, 
as it has been defined by the expectation values of the position operator.
This means that the classical concept of space-time may break down at scales 
where quantum effects become noticeable, and this happens not at Planck 
scales, but already at atomic scales.

There is a wide-spread opinion that the difficulties of a quantum theory of 
gravitation result from the fact that quantum mechanics is defined on 
space-time, while in quantum gravity, this very space-time continuum 
``must be quantized.''
This opinion, obviously, does not make the necessary distinction between 
parameter space-time and physical space-time.

In contrast to parameter space-time, physical space-time has a natural scale.
A scale is, e.g., given by the Bohr radius 
\begin{equation}
\frac{\hbar}{c\, m_e \alpha}
\end{equation}
of the electron in a hydrogen atom.
This scale is determined by the electromagnetic interaction, which in \cite{ws} 
was shown to be a property of the irreducible representations of the 
Poincar\'e group, and by the electron mass $m_e$.
So this mass takes over the role of the (hypothetical) Planck mass in 
characterizing a ``smallest length.''

\section{Geometry of physical\\space-time}

Within an irreducible two-particle representation, the motion of the particles 
relative to each other is determined by a well-defined angular momentum.
The associated Casimir operator $W$ is a constant of the motion.
Quantum mechanics describes this angular momentum in (parameter) space-time
by spherical functions, which in the limit of large quantum numbers describe
probability distributions with the shape of circular orbits.

The circular orbits of a quasi-classical two-particle system, resulting from a 
well-defined angular momentum, can be described by the semi-classical 
expression
\begin{equation}
x_i p_j - x_j p_i = n \hbar   \label{3-0}
\end{equation}
or by
\begin{equation}
p_t = \frac{n \, \hbar}{r} ,          \label{3-0b}
\end{equation}
where $p_t$ is the momentum in the tangential direction.
In words, the tangential momentum is proportional to the curvature 
of the orbit.
Since there are no external forces to keep the particles on these orbits, we are
led to the alternative interpretation that physical space-time, in contrast to 
parameter space-time, has in general a curved metric.

This curvature is not obtained by an active deformation of a predefined 
space-time continuum, but by the {\it ab initio} construction of (physical) 
space-time from (an entangled superposition of) momentum eigenstates within 
an irreducible two-particle representation.
Viewed in this way, it appears more or less trivial that the distribution of 
energy-momentum in space-time should be reflected in the metric of physical 
space-time, and it would be surprising if it were not.

The connection between energy-momentum and space-time is given by the 
factor $e^{-i p x /\hbar}$ in the states (\ref{2-1}).
This factor is invariant under two simultaneous conformal transformations 
\begin{equation}
x \rightarrow \lambda^{-1} x       \label{3-1} 
\end{equation}
and
\begin{equation}
p \rightarrow \lambda \;\; p . \label{3-2}
\end{equation}
By these transformations, not only parameter space-time, but also physical 
space-time, are subjected to a scaling that changes any probability 
distribution in space-time by a scaling factor $\lambda^{-1}$, but keeps the 
form of this distribution invariant.
The symmetry defined by these transformations means that the linear size of 
a structure in space-time is inversely proportional to the magnitude of 
the energy-momentum that defines this structure.
Accordingly, a curvature associated with this structure is directly proportional 
to the energy-momentum.

This especially applies to the curvature of the quasi-classical orbit of two 
particles within an irreducible representation of the Poincar\'e group.
Following Newton's first law, we can describe this orbit as the result 
of a force that acts perpendicularly to the velocity vectors of the particles. 
This force generates a space-like linear momentum perpendicular to their 
actual velocities.
(Remember that the kinematics of the relative momentum is a matter of
a 3-dimensional subspace of $p$ space.)
Such a momentum is described by the momentum flux $T^{ik}, i \not= k\;
(i, k = 1,2,3)$ of the energy-momentum tensor $T^{\mu\nu}$.
The diagonal elements $T^{ii}$ obviously do not contribute to the 
centripetal force. 
Therefore, the deviation of the particles' kinematics from a straight uniform 
motion can, in principle, be deduced from the traceless part of the energy 
momentum tensor.
(Although Lorentz transformations may transform the components of $T^{ik}$ into
$T^{00}$ and $T^{ii}$, these transformations leave the trace of $T^{\mu\nu}$ 
invariant.)
Because the total linear momentum is conserved, the second particle must 
contribute a flux of linear momentum that is opposed to the flux of the 
first.
Metaphorically speaking, both particles exchange momentum.

With this in mind, we now try to express the centripetal forces by a 
non-Euclidean metric of space-time.
Consequently, we have to look for a relation between the curvature of 
space-time and the traceless part of the energy-momentum tensor, as 
the ``cause'' of the curvature.
(Einsteinian gravity, which was set up with the goal of replacing Newtonian gravity, 
uses the trace of the energy-momentum tensor instead. 
Both approaches are in a sense complementary, as far as spherically 
symmetric solutions are concerned \cite{pdm2}.)   
According to what has been said above about conformal scaling, the 
curvature must be proportional to the scaling of the momentum.
Therefore, the curvature experienced by the second particle must be 
proportional to the traceless part of the energy-momentum tensor of the 
first particle, and vice versa.

A curvature tensor that can be set proportional to a traceless energy-momentum 
tensor, must itself be traceless too.
Such a tensor is the Weyl tensor $C^{\mu\nu\sigma\tau}$, which is the traceless 
part of the Riemann curvature tensor $R^{\mu\nu\sigma\tau}$.
From the Weyl tensor, a traceless ``gravitation tensor'' $W^{\mu\nu}$ can be 
derived \cite{pdm}.
This tensor can then be put into relation with the traceless part of the 
energy-momentum tensor $T^{\mu\nu}$.

Examples of traceless energy-momentum tensors, based on different models of 
matter, can be found in \cite{pdm}.
Here we simply subtract the trace from the energy-momentum tensor to make 
it traceless.
This leads to the field equations of {\it conformal gravity} 
\begin{equation}
W^{\mu\nu} = G_{conf} \,(\,T^{\mu\nu} - \frac{1}{2}\, T^\alpha_{\;\; \alpha}
 \, g^{\mu\nu}\,)           \label{3-3}
\end{equation} 
with a ``gravitational constant'' $G_{conf}$.

Conformal gravity has gained interest in recent years because it may solve 
the problems usually associated with ``dark matter'' and ``dark energy'' 
\cite{pdm, pdm2} without additional {\it ad hoc} assumptions. 
Within the scale of our solar system, conformal gravity is known to deliver the 
same results as Einstein's theory of general relativity, which is based on 
the Riemann curvature tensor, rather than on the Weyl tensor \cite{pdm2}.
The problem of ``ghosts,'' which has been encountered in ``quantized'' versions 
of conformal gravity \cite{pdm1, jm}, does not exist for the classical version.

\section{The gravitational constant}

In \cite{ws}, the electromagnetic coupling constant $\alpha$ was calculated 
from the geometry of the parameter space associated with an irreducible 
two-particle state space of spin-1/2 particles.
The same calculation, done for spinless particles, results in a coupling 
constant of $\alpha / 4$.

There is, however, a crucial difference between quantum electrodynamics
and gravitation.
Whereas in quantum electrodynamics it makes sense to consider an isolated
two-particle system, this is an unrealistic configuration in gravitation.
There is no way to set up a ``neutral'' environment or to ``shield'' gravitation.
Therefore, an experimental setup for a ``scattering experiment'' in analogy to
electron--electron scattering must always take into account the whole 
environment.
This means we have to take into account at least $10^{80}$ heavy particles,
which is the estimated number of protons in the (observable) universe 
\cite{ou}.

A gravitational scattering experiment of an (electrically neutral) particle 
of, say, the mass of the proton, includes at first the {\it selection} of a second 
particle from $10^{80}$ available particles.
This is followed by a {\it transition} from the ``incoming'' two-particle pure 
product state to an irreducible (entangled) two-particle state.
Finally, we have a {\it transition} to an ``outgoing'' two-particle pure product 
state, which is the quantum mechanical description of measuring the individual 
momenta of the particles after the scattering has taken place.
Note that there are two transitions between pure product states and 
entangled state, but only one selection.

The following is an attempt to quantum mechanically describe the ``selection  
process.''
The selection of a partner particle will be considered as a ``transition'' from
an ``incoming'' one-particle state (of the first particle) to a two-particle state.
For the first particle, there are $10^{80}$ independent ways to form a 
two-particle state.
Let us describe the corresponding quantum mechanical transition amplitude by a 
state in a $10^{80}$-dimensional state space.
Then the states of this state space have to be normalized by the factor 
$1/10^{40}$.

This normalization ensures that the total transition probability from a 
specific incoming (one-particle) state to an outgoing (one-particle) 
state, through any intermediate two-particle state, equals unity.
On the other hand, the field equations (\ref{3-3}) describe the contribution 
of only a specific second particle, characterized by its 
energy-momentum tensor at a point $x$, to the curvature of space-time.
Accordingly, the scattering process contains only the transitions up to the 
outgoing two-particle product state.
For this reason, the ``selection amplitude'' enters only once.
The normalization factor in this amplitude leads to an additional factor of 
$1/10^{40}$ to the two-particle coupling constant $\alpha / 4$.
The resulting ``gravitational coupling constant''  is then in good agreement 
with the empirical strength of the gravitational interaction, which is known 
to be weaker than the electromagnetic interaction by a factor of $10^{40}$.

\section{``Quantum gravity''}

The field equations (\ref{3-3}) describe a classical theory of gravitation.
What, then, is their quantum mechanical analogue?
Since we just have sketched a connection between quantum theory and 
classical conformal gravity, we are able to give an answer to this 
question:
The quantum mechanical basis of conformal gravity is nothing other than 
an irreducible two-particle representation of the Poincar\'e group. 
In other words, there is no specific ``quantum gravity'' apart from 
the common rules of relativistic quantum mechanics.
The situation is similar to quantum electrodynamics, as discussed in 
\cite{ws}: Gravity emerges from the restrictions on the two-particle 
state space imposed by the condition of irreducibility.

\section{Conclusions}

Reasons have been given as to why gravitation can be understood as a 
basic property of relativistic quantum mechanics, more precisely, as a
property of the irreducible two-particle representations of the Poincar\'e group. 
Gravitation is not provided by ``coupling'' to an ``external field.''
Rather it is the outcome of correlations within the quantum mechanical 
state-space of matter resulting from the condition of irreducibility.  
These correlations lead to the equations of classical conformal gravity.
In short, gravitation is a quantum mechanical property of matter.

Physical space-time turns out to be just another quantum mechanical property 
of matter.
Its geometry in the large is determined by the equations of conformal gravity.
Its scale in the small is defined by the electromagnetic interaction and by the 
masses of the particles involved in this interaction.
Together, the electromagnetic and gravitational interactions provide the basis for 
building extended atoms, molecules, and macroscopic bodies, to fill up 
space-time.
The electromagnetic interaction provides photons, which can be used to unveil 
the geometry of space-time to an observer.
Needless to say, the electromagnetic interaction establishes a causal structure 
in space-time.
It is these interactions that make the difference between parameter 
space-time and physical space-time.
Therefore, the emergence of physical space-time goes in parallel with the 
emergence of interactions.

The validity of classical space-time ends at scales where quantum 
mechanics becomes effective. 
These scales are related to the electron mass, rather than to the Planck mass.
There is no room for the latter, because it is not possible to construct
a mass from $\hbar$, $c$, and $G_{conf}$.

\newpage  % to adjust both columns


\begin{thebibliography}{99}


\bibitem{ws} W. Smilga, ``Reverse Engineering Approach to Quantum Electrodynamics,''
{\it J. Mod. Phys.}, Vol. 4, No. 3, 2013, in press.\\
\textcolor{blue}{\href{http://arxiv.org/abs/1004.0820}{arXiv:1004.0820}} 

\bibitem{sss} S. S. Schweber, 
``An Introduction to Relativistic Quantum Field Theory,'' pp. 44--46,
Harper \& Row, New York, 1962. 
 
\bibitem{are} A. R. Edmonds, ``Angular Momentum in Quantum Mechanics,'' pp. 27--29,
Princeton University Press, Princeton, 1957.

\bibitem{in} I. Newton, ``Philosophi{\ae} Naturalis Principia Mathematica,''
London, 1687.

\bibitem{nw} T. D. Newton and E. P. Wigner, ``Localized States for Elementary Systems,'' 
{\it Rev. Mod. Phys.}, Vol. 21, 1949, pp. 400--406.

\bibitem{rhbook} R. Haag, ``Local Quantum Physics,'' pp. 31--33,
Springer-Verlag, Berlin, 1996.

\bibitem{pdm} P. D. Mannheim, ``Alternatives to Dark Matter and Dark Energy,''
{\it Prog. Part. Nucl. Phys.}, Vol. 56, 2006, pp. 340--445.\\
\textcolor{blue}{\href{http://arxiv.org/abs/astro-ph/0505266}{arXiv:astro-ph/0505266}}

\bibitem{pdm2} P. D. Mannheim, ``Making the Case for Conformal Gravity.''\\
\textcolor{blue}{\href{http://arxiv.org/abs/1101.2186}{arXiv:1101.2186}}

\bibitem{pdm1} C. M. Bender and P. D. Mannheim,
``No-Ghost Theorem for the Fourth-Order Derivative Pais--Uhlenbeck Oscillator 
Model.''\\
\textcolor{blue}{\href{http://arxiv.org/abs/0706.0207}{arXiv:0706.0207}}

\bibitem{jm} J. Maldacena, ``Einstein Gravity from Conformal Gravity.''\\
\textcolor{blue}{\href{http://arxiv.org/abs/1105.5632}{arXiv:105.5632}}

\bibitem{ou} Wikipedia: ``Observable Universe.''\\
\textcolor{blue}{\href{http://en.wikipedia.org/wiki/Observable\_universe}{wikipedia:Observable universe}}

\end{thebibliography}
\end{document}